\documentclass[runningheads]{llncs}

\usepackage{amsmath}
\usepackage{amsfonts}
\usepackage[numbers]{natbib}
\usepackage{booktabs}
\usepackage{graphicx}

\begin{document}

\title{Mapping Between fMRI Responses to Movies and their Natural Language Annotations}

\titlerunning{Mapping Between fMRI Responses and Semantic Representations}

\author{Kiran Vodrahalli${}^1$ \and Po-Hsuan Chen${}^1$ \and Yingyu Liang${}^1$ \and \\
Christopher Baldassano${}^1$ \and Janice Chen${}^2$ \and Esther Yong${}^3$ \and Christopher Honey${}^2$ \and \\
Uri Hasson${}^1$ \and Peter Ramadge${}^1$ \and Kenneth A. Norman${}^1$ \and Sanjeev Arora${}^1$}

\authorrunning{Kiran Vodrahalli et al.}

\institute{
${}^1$Princeton University, ${}^2$Johns Hopkins University, ${}^3$University of Toronto}

\maketitle

% no jargon; 
\vspace{-2em}
\begin{abstract}

Several research groups have shown how to correlate fMRI responses to the meanings of presented stimuli. This paper presents new methods for doing so when only a natural language annotation is available as the description of the stimulus. We study fMRI data gathered from subjects watching an episode of BBC’s Sherlock \cite{Chen17}, and learn bidirectional mappings between fMRI responses and natural language representations. We show how to leverage data from multiple subjects watching the same movie to improve the accuracy of the mappings, allowing us to succeed at a scene classification task with $72\%$ accuracy (random guessing would give $4\%$) and at a scene ranking task with average rank in the top $4\%$ (random guessing would give $50\%$). The key ingredients are (a) the use of the Shared Response Model (SRM) and its variant SRM-ICA \cite{Chen15, Zhang16} to aggregate fMRI data from multiple subjects, both of which are shown to be superior to standard PCA in producing low-dimensional representations for the tasks in this paper; (b) a sentence embedding technique adapted from the natural language processing (NLP) literature \cite{Arora17} that produces semantic vector representation of the annotations; (c) using previous timestep information in the featurization of the predictor data.

\end{abstract}

\vspace{-3em}
\section{Introduction}
\vspace{-0.5em}

Recent work has provided convincing evidence that fMRI readings from human subjects can be related to semantics of presented stimuli. Such experiments consist of finding $(1)$ low-dimensional representations of the fMRI signals, and $(2)$ low-dimensional semantic representations of the external stimulus. These tasks often build upon work in machine learning. 

The earliest work concerned simple settings with carefully controlled stimuli, such as subjects being presented (visually or auditorily) with one of a set of carefully selected words \cite{Mitchell08}. The semantic representation of a word was computed using word embeddings, a tool from natural language processing \cite{Deerwester90} that represents each word as a point in a k-dimensional meaning space. This work was extended \cite{Pereira11, Pereira16} to perform ``brain reading'', using fMRI readings and a popular text-analysis tool called topic modeling to reconstruct word clouds from brain activity evoked by a word/concept stimulus. 

The next obvious step in this research program is to understand fMRI readings collected from subjects as they process more complex stimuli such as movies. In such settings it is not clear how to represent the semantics of the stimulus, since a multitude of signals (auditory as well as visual) are presented within a short time interval. Ideally, this mapping between fMRI and stimuli should be meaningful across different human subjects, so that the accuracy of matching the two should improve by using data from multiple subjects. One approach to solving this task was presented in \cite{Huth12}, which studied fMRI responses to a natural movie stimulus. In this case, the movie stimulus was represented with a feature space of $1705$ distinct nouns and verbs. A subsequent study \cite{Huth16} examined fMRI responses to audio stories, and departed from the previous work by applying distributional embeddings to featurize the dialog and predict voxel activation. The goal in these papers was to derive a semantic word map for the voxels of the brain. Another paper \cite{Wehbe14} gathered fMRI data from subjects reading a story, and used unweighted averages of distributional embeddings to featurize sentences for predicting voxel activity.

In this paper, we study the Sherlock fMRI dataset \cite{Chen17}, which consists of fMRI recordings of $16$ people watching the British television program ``Sherlock'' for $50$ minutes broken into $1973$ TRs, where each TR is $1.5$ seconds of film. As a proxy for the semantics of the movie, we use externally annotated English text scene annotations of the program (average annotation length $15$ words per TR). We examine brain data from predefined regions of interest (ROIs) in the brain, and separately analyze each one. In particular, we examine the default mode network (DMN), dorsal and ventral language areas, the occipital lobe, and a $26000$-voxel mask containing voxels with high intersubject correlation across the whole brain.
We seek to determine whether various modifications to fMRI and text featurization as well as the usage of previous timepoint information help to improve bidirectional mappings between fMRI data and semantic meaning vectors. In particular, we examine the effects of three featurization methods for fMRI and text data: \textit{Low-dimensional shared fMRI representation} across subjects, \textit{weighted semantic embeddings} of text annotations, and using \textit{previous timepoints} in the performance of linear maps between people.

\vspace{0.5em}

\textbf{Aggregating fMRI responses across subjects}. In prior work, combining fMRI response data from multiple subjects is often solved by averaging, anatomical alignment and smoothing, or latent multivariate feature modeling \cite{Wehbe14, Conroy13, Huth16}. Further work concludes that high-level representations of content from movies are shared across people and that there can be considerable de-noising benefits from averaging across people \cite{Chen17}. Another recent paper \cite{Chen15} introduced the Shared Response Model (SRM), an algorithm that stems from previous work on hyperalignment \cite{Haxby11}. The SRM in \cite{Chen15} optimizes the objective $\sum_{i = 1}^n \|X_i - W_iS\|_F$ for a low-dimensional shared space $S$ and orthogonal-column subject specific maps $W_i$, and can be thought of as a multi-subject extension of PCA. Simultaneously reducing dimensionality across subjects outperforms other averaging approaches at matching up specific timepoints in a movie across subjects. 

\vspace{0.5em}

\textbf{Semantic representation of stimulus}. To find semantic representations of English annotations, it is natural to draw upon related work in natural language processing. One common approach involves word embeddings created by using co-occurrence information in a large corpus like Wikipedia. A simple technique for representing longer pieces of text is to average the vectors for the individual words \cite{Wehbe14}. Recently, this simplistic idea has been extended in natural language processing by using recurrent neural nets \cite{Kiros15} or by modifying the original model for learning word vectors to learn word sequence chunks (for instance, paragraphs) directly from the text \cite{Le14}. These more powerful methods have the drawback of requiring large corpora, making them unusable in our current setting where we only have $1973$ brief text annotations. Very recently, \cite{Arora17} suggested a simpler method for this task that requires no additional information beyond the existing word embeddings, yet beats these more complicated methods in standard natural language tasks. We adapt this method to construct \textit{annotation embeddings} using weighted combinations of the vector representations for the words in each annotation. One of our key results is that this new embedding significantly outperforms unweighted averaging of word vectors. 

\vspace{0.5em}

\textbf{Using previous timestep information}. A movie stimulus naturally breaks up into multi-timestep scenes that occur at different timepoints. Thus, at any given timepoint, there may be a window of previous timesteps that are part of the current scene and thus are relevant to understanding the current time point in both fMRI and Text space. We would like to incorporate this past information shared within scenes in order to learn better maps between fMRI and Text. Other models \cite{Huth16, Wehbe14} incorporate past information by modeling the hemodynamic response function (HRF) that describes the fMRI BOLD response to a stimulus. However, this approach focuses on small timescales, and only accounts for the delayed and temporally-smeared BOLD response rather than attempting to aggregate scene information. Our approach is to first approximate the HRF delay with a simple one-time shift of $4.5$ seconds, and to then incorporate longer time-scales into our model by including in the featurization a $k$-sized window of previous timesteps, where $k$ is varied from $0$ to $30$ (these numbers correspond to $0-45$ seconds). 

\vspace{0.5em}

\begin{figure}[!t]
\centering
\includegraphics[width=1\textwidth]{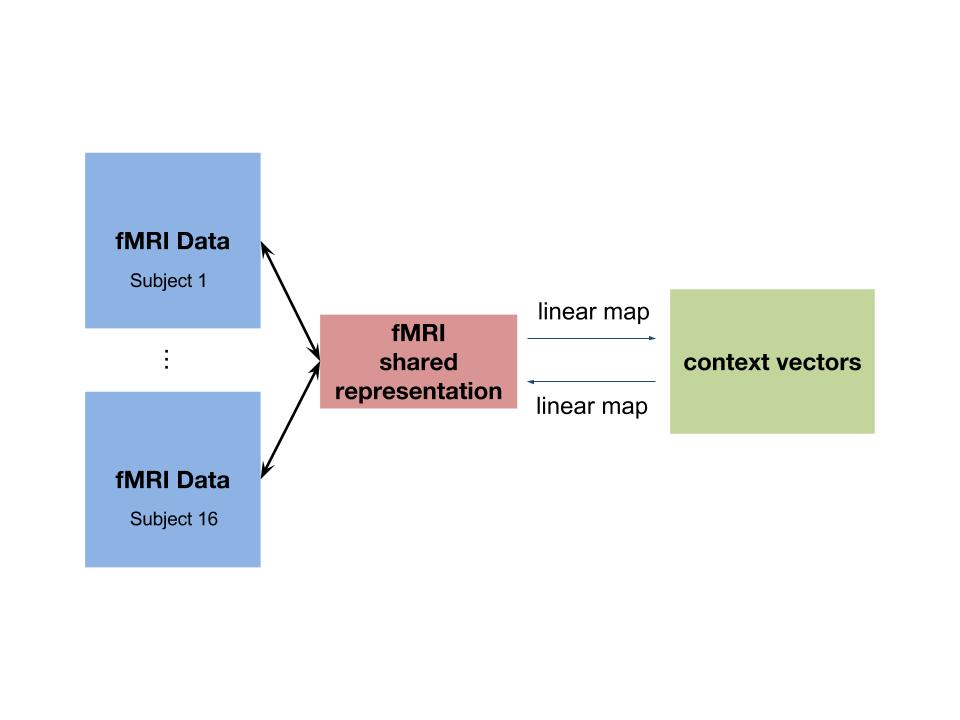}
\caption{Summary of Experimental Setup: We learn a shared response for the brain activity of $16$ different subjects watching BBC's Sherlock, construct semantic featurizations for associated semantic annotations, and learn bidirectional linear maps between the two data modes.}
\label{fig:setup}
\end{figure}

To evaluate the effect of each of these featurization methods, we use linear maps to relate the fMRI signal to the representation of the semantic content, using only the first half of the movie. These maps are validated with two experiments: scene classification and scene ranking.  We divide up the second half of the movie into $25$ uniformly-sized chunks. \textit{Scene classification} is the task of using correlation to match predicted intervals of fMRI or semantic activity with the ground truth, and reporting the percentage of the time that the match is perfect. Since there are $25$ intervals, random chance performance at this task is $4\%$. \textit{Scene ranking} is the same task, except we measure the average rank of the correct answer: Random chance performance here is $50\%$. For a visual summary of the setup, see Figure \ref{fig:setup}. These experiments are executed with the fMRI $\to$ Text maps (given fMRI data, predict text annotations) as well as the Text $\to$ fMRI maps (give text annotations, predict fMRI data).

\subsection{Main results}

Our main results are (i) showing that fMRI responses from multiple individuals can be effectively combined using SRM to improve the matching accuracy ($1.3\times$ average improvement over our baseline, the average PCA representation) between the fMRI and the text annotation (Table \ref{table:avg_deltas}, Figures \ref{figure:FT4}, \ref{figure:TF4}), (ii) demonstrating that a method for combining word vectors into annotation vectors via a suitable weighting \cite{Arora17} for averaging word vectors on average improves $1.2\times$ over unweighted averaging (Table \ref{table:avg_deltas}, Figures \ref{figure:FT4}, \ref{figure:TF4}), and (iii) finding that appropriate inclusion of information from previous time steps yields as much as a $5.3\times$ improvement (on average, $1.8\times$) in tasks measuring the performance of mapping from fMRI to Text (see Figure \ref{figure:FT4}, Dorsal Language ROI). There are diminishing returns after a certain point to including more time steps: The optimal number seems to be around $5-8$ previous time steps. For the Text $\to$ fMRI task, using previous time steps decreases performance.

We also report the top performances for each task. For the fMRI $\to$ Text task, our top scene classification performance is $72\%$ accuracy, meaning that for $72\%$ of the time intervals we examine, our predicted annotation representation correlates the most with the true annotation representation for that time interval (see Figure \ref{figure:topROIacc}, Whole Brain ROI). Notably, this result improves considerably over the random guessing rate of $4\%$. The corresponding scene ranking performance is $96\%$, meaning that on average, the rank of the true annotation representation is within the top $4\%$ when sorted by correlation with the predicted annotation representation. The Text $\to$ fMRI task had worse results. The top scene classification performance for Text $\to$ fMRI is $56\%$ accuracy, and the corresponding scene ranking accuracy is $91\%$ (see Figure \ref{figure:topROIacc}, DMN-A ROI).

\vspace{-0.5em}
\section{Methods}
\vspace{-0.5em}

\subsection{Preprocessing the Dataset}

\begin{figure}[!t]
\centering
\includegraphics[width=0.5\textwidth]{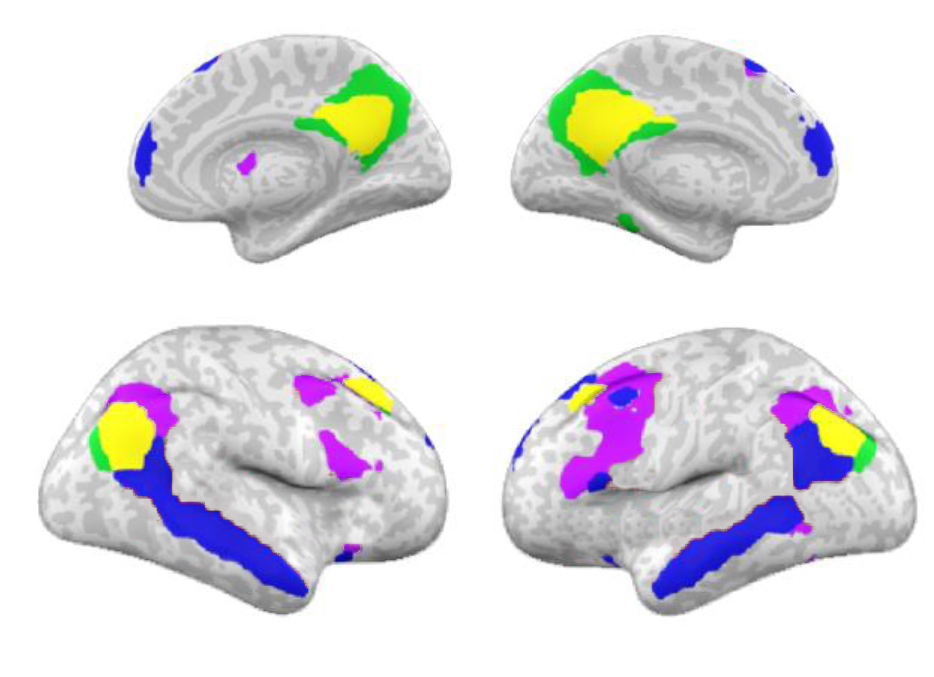}
\includegraphics[width=0.2\textwidth]{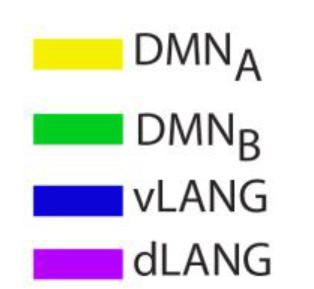}
\caption{Visualization of the DMN and Ventral/Dorsal Language Area ROIs \cite{Simony16}: Here, we display four of the regions of interest on a brain map. These masks were collected on the Pie Man dataset \cite{Simony16}, then fit to a standard anatomical brain (MNI$152$), and interpolated to $3$-mm isotropic voxels \cite{Simony16}. In order to define the DMN-A and DMN-B regions, as well as the Ventral and Dorsal language area regions, the intersubject functional correlation matrix \cite{Hasson04} was calculated from the fMRI data of $36$ subjects collected while they were listening to stories \cite{Simony16}. Then, $k$-means clustering was applied to find the networks. The DMN-A and DMN-B networks were identified by comparing the resultant clusters to the DMN ROIs derived via thresholding the functional correlation between the posterior cingulate (identified by literature) and the rest of the brain for the fMRI data of $36$ subjects during resting state \cite{Simony16}. The Ventral and Dorsal language areas were identified by comparing the clusters to previous results in the literature \cite{Simony16}.}
\label{fig:ROIvis}
\end{figure}

Before performing any analysis, the fMRI data are preprocessed and standardized using the techniques described in \cite{Chen17}. Then, we identify six distinct brain regions of interest (ROIs) that we treat completely separately. That is, we first apply ROI masks to the whole-brain data and then learn SRM-representations for each of these ROIs separately. We use the ROIs for the default mode network (DMN-A, DMN-B) and the ROIs for the ventral and dorsal language areas identified in \cite{Simony16}. Methodology for finding the default mode network relies on intersubject functional correlation (ISFC), a technique first introduced by \cite{Hasson04}. The central idea is that natural stimuli (like movies) evoke reliable, time-dependent activity across a variety of brain networks. For more details, see Figure \ref{fig:ROIvis}. We are interested in the DMN ROIs in particular since prior work has demonstrated that these regions play a crucial role in tracking the narrative in settings such as watching movies or reading stories \cite{Hasson04, Hasson10, Honey12, Regev13, Ames15, Simony16, Yeshurun17}.
The ``Whole Brain'' ROI is a $26000$-voxel mask of the brain that highlights voxels that have intersubject correlation $> 0.2$ on the data, and the Occipital Lobe ROI is defined from the MNI Structural Atlas in FSL (\texttt{https://fsl.fmrib.ox.ac.uk/fsl/fslwiki/Atlases}). We include these ROIs for holistic comparison across the whole brain.

We also truncate the first three TRs of fMRI data and the last three TRs of semantic annotation data. This operation effectively aligns the fMRI and semantic data under the assumption that there is a $4.5$ second delay between the onset of the stimulus and the BOLD response signal. 

\subsection{Constructing and Aggregating Semantic Vectors}
\vspace{-0.5em}
In order to represent words, we take advantage of the distributional properties of words in a large corpus - namely, English Wikipedia. We train word embeddings as described in \cite{Arora16}, which perform on par with other standard word embedding techniques like GloVe and Word$2$Vec \cite{Arora16}. Now, we diverge from the prior work by calculating and applying a domain specific re-centering of the embeddings. After creating an embedding for each word in the vocabulary of the Sherlock annotations, we calculate the top principal component of all word embeddings in the vocabulary. We then scale the normalized top principal component by the average Euclidean norm of a word embedding in the Sherlock vocabulary. This vector represents a kind of average topic for the Sherlock vocabulary. Since we would like our word embeddings to be discriminative within this average topic, we algebraically subtract out this component. We can view this step as finding a translation operation that moves the word embeddings away from the region of semantic space that is close to generic words in the Sherlock annotation corpus. 

The central assumption in \cite{Arora16} is the probability model for a word $w$ in a vocabulary $V$ given a context $c$, where the context represents a small window of words in the corpus. This model is given by $\textstyle \mathcal{P}\left[w | c\right] = \frac{1}{Z_c}\textnormal{exp}(v_w^Tc)$ where $v_w$ represents the vector for a given word and $Z_c$ is a term that normalizes the distribution. The idea is that the context vector $c$ represents the subject matter of the text at a given point in time. 

Using this assumption and a few others, the word vector learning problem is phrased in \cite{Arora16} as the \textit{squared-norm objective}:  
\[
\textstyle \min_{\{v_w\}_{w \in V}, C} \sum_{w_1, w_2} X_{w_1, w_2}\left(\log(X_{w_1, w_2}) - \|v_{w_1} + v_{w_2}\|_2^2 - C\right)^2
\]
where $C$ is a bias term, $X$ is the co-occurrence count matrix between single words in a small window of text (fixed at $\approx 5$ words) and $v_w$ are the word vectors we are trying to learn. This objective can be optimized with gradient descent. For a full treatment of the theoretical properties of the word vectors and the derivation of the squared-norm objective, see \cite{Arora16}.

For every $1.5$-second time-point in our Sherlock movie, annotators were asked to provide a natural description of what is happening in the movie: actions, dialog, and so on. This annotation is typically a few sentences long, and contains around $15$ words on average. We can think of each annotation as the current context of the movie narrative. The log-linear probability model of \cite{Arora16} for words given context $c$ implies that the maximum likelihood estimator of the context is simply the average of all words in the annotation. (This formulation is a theoretical justification for a standard rule of thumb in natural language processing for representing the sense of a small piece of text by the average of the embeddings for the words in the text). We will call these representations the \textbf{unweighted} annotation vectors. 

However, one imagines that not all words in the annotation are equally important, and that a better representation might be possible by taking this idea into account. This approach has been studied in various neural network frameworks \cite{Kiros15}; however, applying these kinds of models requires a large annotation corpus, while we only have $1973$ $15$-word annotations. A recent paper \cite{Arora17} suggests a principled approach for computing a representation of a small piece of text. The intuition from \cite{Arora17} is that words that occur with much greater frequency in the original corpus may inherently contain less information, since these words are in some sense uniform with respect to the whole word distribution. Therefore, more frequent words should be weighted less. The paper \cite{Arora17} modifies the above language generation model as follows: For a word $w$ given context $c$, the probability of a word $w$ given context $c$ is 
\begin{align}
\textstyle \mathcal{P}\left[w | c\right] = \alpha \mathcal{P}\left[w\right] + (1 - \alpha)\frac{\textnormal{exp}(v_w^Tc)}{Z_c}
\end{align}
where $Z_c$ normalizes the distribution and $\alpha \in [0, 1]$. We can think of this model as a weighted sum of the probability of a word $w$ appearing not conditioned on the context $c$ and the probability of a word $w$ appearing conditioned on the context $c$.

\begin{figure}[!t]
\centering
\includegraphics[width=1\textwidth]{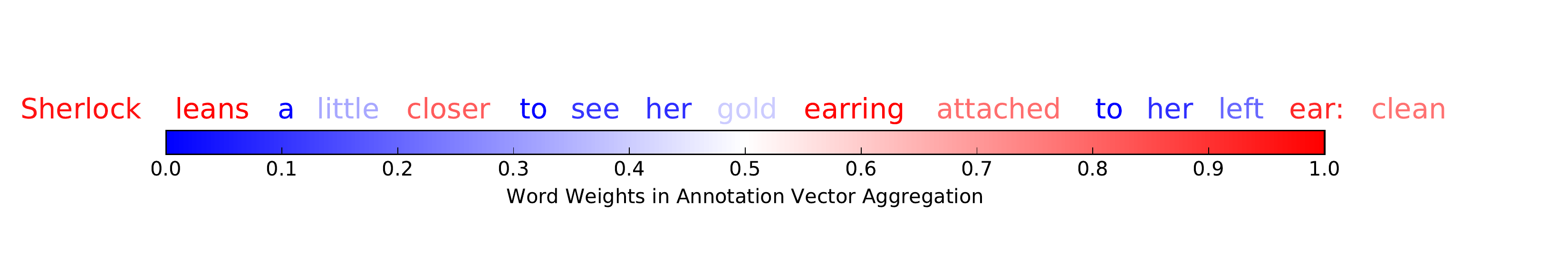}
\vspace{-2.5em}
\caption{Visualization of Semantic Annotation Vector Weightings: We display an example sentence from the Sherlock annotations, where we have colored important words red, and unimportant words blue. Brighter red means more important, and darker blue means less important.}
\label{fig:semantic_weighting}
\vspace{-1em}
\end{figure}

The revised estimate of the context vector $c$ in this modified objective is
\begin{align}
v_{\textnormal{annotation}} = \sum_{\textnormal{word} \in \textnormal{annotation}} \frac{\beta}{\beta + p_{\textnormal{word}}} \cdot v_{\textnormal{word}}
\end{align}
where $\beta := \frac{1- \alpha}{\alpha Z}$. Typically, we choose $\alpha$ such that $\beta \approx 10^{-4}$.
These representations are called the \textbf{smooth inverse frequency (SIF)} annotation vectors, or \textbf{weighted} annotation vectors. Figure \ref{fig:semantic_weighting} depicts a example sentence with the respective word weights colored according to importance in the sentence embedding.

Using either the unweighted or weighted approach will produce one annotation vector for each of our $T$ time steps. On the training portion of the data (the first half of the movie), we calculate an average annotation vector and subtract it from all data. Here, we assume that the average annotation vector is invariant, which turns out to be a good assumption.

\vspace{-0.5em}
\subsection{Shared Response Models for Multi-Subject fMRI}
\vspace{-0.5em}
The Shared Response Model (SRM) \cite{Chen15} is an unsupervised probabilistic latent variable model for multi-subject fMRI data under a time-synchronized stimulus. From each subject's fMRI view of the movie, SRM learns projections to a shared space that captures semantic aspects of the fMRI response.

Specifically, SRM learns $N$ maps $W_i$ with orthogonal columns such that $\|X_i - W_i S\|_F$ is minimized over $\{W_i\}_{i = 1}^N, S$, where $X_i\in \mathbb{R}^{v \times T}$ is the $i^{th}$ subject's fMRI response ($v$ voxels by $T$ repetition times) and $S\in \mathbb{R}^{k\times T}$ is a feature time-series in a $k$-dimensional shared space. In this paper, $k = 20$ since low-rank SVD with $20$ dimensions captures $90\%$ of the variance of the original fMRI matrices \cite{Chen17}. We also experimented with using $k = 50, 80, 100, 1000$, but the results barely varied from using $k = 20$ dimensions.
Note that, for testing, the learned $W_i$ allow us to project unseen fMRI data into the shared space via $W_i^TX_i^{\textnormal{test}}$ since $W_i$ has orthogonal columns.

We also examine a variant of SRM called SRM-ICA \cite{Zhang16} that modifies the SRM algorithm with an independent components analysis (ICA) objective. ICA is an unsupervised learning technique that identifies independent signals from a mixture by looking for rotations of the data that produce non-Gaussian signals. SRM-ICA brings this approach to learning a shared space: While in SRM we alternated by solving for $W_i$ by minimizing $\|X_i - W_i S\|_F$ and updating $S$ with the average of $W_i^TX_i$, we change the objective we use to update each $W_i$ to an ICA objective: Maximizing the non-Gaussianity of the shared response $S = \frac{1}{n} \sum_{i = 1}^n W_i^+X_i$, individually with respect to each $(X_i, W_i)$ pair.

In our experiments, we compare average SRM and SRM-ICA projections ($\frac{1}{N}\sum_{i = 1}^N W_i^TX_i^{\textnormal{test}}$) against the baseline average principal components analysis (PCA) projections. PCA is a standard linear dimensionality reduction technique that finds an optimal (in Frobenius norm) orthogonal projection of the data onto a low-dimensional subspace.

\vspace{-1.5em}
\subsection{Learning Linear Maps}
\vspace{-0.5em}

Our approach to predicting semantic annotation vectors from fMRI vectors and vice versa is simply linear regression with two kinds of regularization. Letting $X \in \mathbb{R}^{v \times T}$ represent the fMRI data matrix (either SRM, SRM-ICA, or PCA) for a specific ROI and $Y \in \mathbb{R}^{100 \times T}$ represent the annotation vectors, our main approach is given by solving the Procrustes problem $\min_{\Omega} \|Y - \Omega X\|_2^2 $ with orthogonal columns constraint $\Omega^T\Omega = I_{v \times v}$. Thus, we learn a matrix
$\Omega \in \mathbb{R}^{100 \times v}$ as a map from $X \to Y$, decoding fMRI vectors into semantic space. 
Our other approach is given by the ridge regression problem $\min_{\omega_j} \|y_j - \omega_j^TX\|_2^2 + \|\omega_j\|_2^2 $
where $j \in [1, 100]$ for each word vector dimension. Putting the $\omega_j$ together forms $\Omega \in \mathbb{R}^{100 \times v}$ as before, with the orthogonality constraint replaced by a row-wise $\ell_2$-norm regularization.

\vspace{-1.5em}
\subsection{Adding Previous Timesteps}

\begin{figure}[!t]
\centering
\includegraphics[width=1\textwidth]{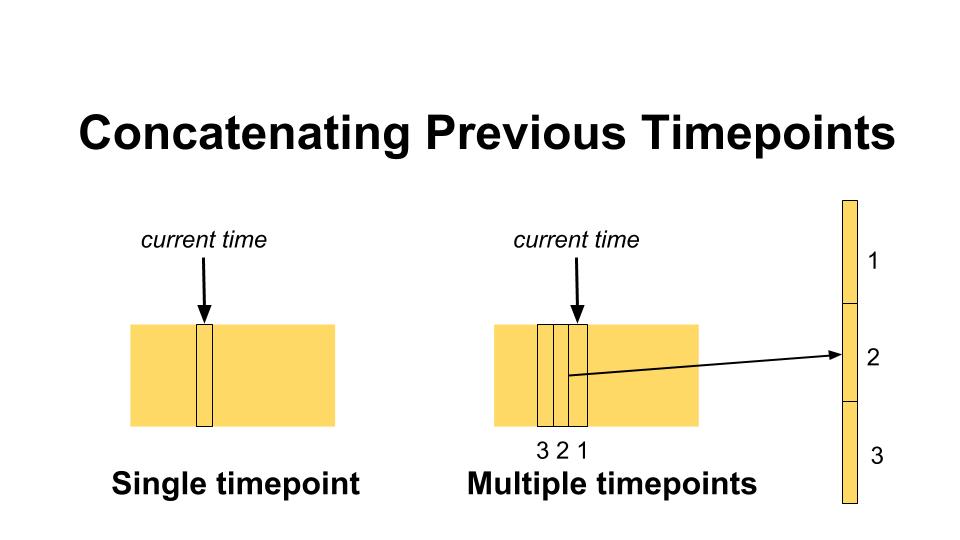}
\caption{Visualizing Concatenation: We visualize what the single timestep case looks like compared to a case where we use the previous two timesteps in our featurization as well. The latter case results in a more complicated model, since one of the dimensions of our linear map triples in size.}
\label{figure:vis_prev_times}
\end{figure}

One could augment the fMRI and annotation vectors using past time steps by finding a complicated combination of the features at each time step, resulting in a representation with the same number of dimensions. For now, we sidestep the complexity of this task by simply concatenating $k$ previous vectors to the predictor vector at each time step (TR) before learning mappings as before. A potential downside to this approach is that we linearly increase the dimensionality with $k$, which can be intractable for large $k$. However, this approach allows every predictor feature at every timepoint to have its own weight in the linear map, creating a powerful model. Thus, in the fMRI $\to$ Text case, we stacked the $k$ previous fMRI vectors onto each fMRI vector, and did not modify the textual annotation vectors. In the Text $\to$ fMRI case, we stacked $k$ previous text annotation vectors and left the fMRI vectors unmodified. When previous time steps do not exist, we append an all-zeros vector instead. We can think of the modified representations as capturing a notion of the dynamics occurring over an interval of $1.5(k+1)$ (TR length $\times$ total number time points) seconds. In this paper, we tried $k = 1$ to $9$ in steps of $1$, and then $k = 10$ to $30$ in steps of $5$. See Figure \ref{figure:vis_prev_times} for a visualization.  

\vspace{-1.5em}
\subsection{Experiment Descriptions}

First, we divide our $1973$ TRs into $50$ uniformly-sized chunks of time, the first $25$ of which are our training data and the latter $25$ of which are our testing data. We learn maps both from fMRI to text annotations and from text annotations to fMRI on the training data. From now on, we refer to \textbf{fMRI $\to$ Text} experiments as those which take an fMRI representation as input and attempt to predict a semantic annotation vector representation. Likewise, \textbf{Text $\to$ fMRI} experiments are those which take in a semantic annotation vector input and predict an fMRI representation. Also note that we train the linear maps on the individual TRs as opposed to the $25$ chunks. 

We perform two primary experiments in this paper, \textbf{scene classification} and \textbf{scene ranking}. These experiments are applied to both the fMRI $\to$ Text and Text $\to$ fMRI settings. In the following description, we denote the predictor space by $X$ and the target space by $Y$.  

Suppose we are in the $X \to Y$ setting. For each time chunk $i \in [1, 25]$ in $X$-space, we predict chunk $i$ in $Y$-space using the learned map, by applying the map individually to each TR within the time chunk. Then, we calculate the Pearson correlation of the predicted chunk $i$ (represented by concatenating the representations for each TR in the chunk into one long vector) with each of the actual time chunks $j \in [1, 25]$, and we rank the chunk indexes by correlation. 

\textbf{Scene classification}. Given the ranking of actual time chunks by correlation with the predicted chunk, we report the proportion of the time that the correct chunk index is ranked the highest. This measure has a $4\%$ chance rate, meaning that if we randomly ranked the actual chunks, any particular chunk would be the top chunk $4\%$ of the time.

\textbf{Scene ranking}. Given the ranking of actual time chunks by correlation with the predicted chunk, we calculate $1 - \frac{\textnormal{average rank of the correct index}}{25}$. This measure has $50\%$ chance rate, meaning that if we randomly ranked the actual time chunks, the average rank of any particular chunk would be in the middle.

We report both of these metrics because the $4\%$ chance rate task gives a better idea of the distribution of the ranking, while other authors have used the $50\%$ chance rate, obtaining ranking scores between $70\% - 80\%$ \cite{Pereira11, Wehbe14, Pereira16}.

We also give some additional analysis of the properties of stacking previous time points, and discuss how they affect prediction capabilities. In particular, we observe the dependence of classification accuracy on the number of previous time steps.

\vspace{-1em}
\section{Results}

\begin{table}[!t]
\setlength\tabcolsep{0.9em} 
\begin{tabular}[t]{l | c | c }
fMRI $\to$ Text & \begin{tabular}{@{}c@{}} Maximum \end{tabular} 
& \begin{tabular}{@{}c@{}} Average \end{tabular} \\
\midrule
Previous Timesteps vs. None & $5.3\times$ & $1.8\times$  \\
Procrustes vs. Ridge & $2.8\times$ & $1.3\times$  \\
SRM/SRM-ICA vs. PCA & $1.8\times$ & $1.3\times$ \\
Weighted-SIF vs. Unweighted & $1.6\times$ & $1.2\times$ \\
\midrule
\midrule
Text $\to$ fMRI & \begin{tabular}{@{}c@{}} Maximum \end{tabular}
& \begin{tabular}{@{}c@{}} Average \end{tabular} \\
\midrule
Previous Timesteps vs. None & $2.5\times$ & $0.5\times$ \\
Procrustes vs. Ridge & $3.0\times$ & $0.8\times$ \\
SRM/SRM-ICA vs. PCA & $2.3\times$ & $1.2\times$ \\
Weighted-SIF vs. Unweighted & $1.8\times$ & $1.1\times$ \\
\end{tabular}
\vspace{1em}
\centering
\caption{Table of Improvement Ratios for Various Algorithmic Parameters: In this table we give the maximum and average improvement ratios for a specific algorithmic technique over another, including usage of previous time steps, SRM/SRM-ICA versus PCA, SIF-weighted annotation embeddings versus unweighted annotation embeddings, and Procrustes versus ridge regression for both fMRI $\to$ Text and Text $\to$ fMRI. When we use previous timesteps, we consider the results for using $5-8$ previous time steps. These numbers are all for the scene classification task. Note that the values from the maximum columns can be seen visually in Figures \ref{figure:FT4} and \ref{figure:TF4} respectively.} \label{table:avg_deltas}
\end{table}

\begin{figure}[!t]
\centering
\includegraphics[width=1\textwidth]{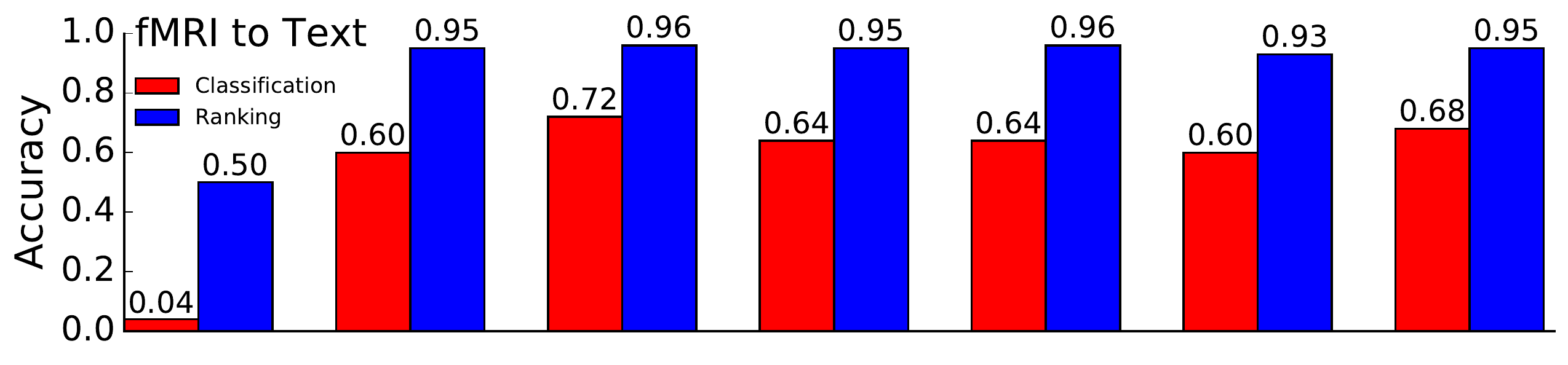}
\includegraphics[width=1\textwidth]{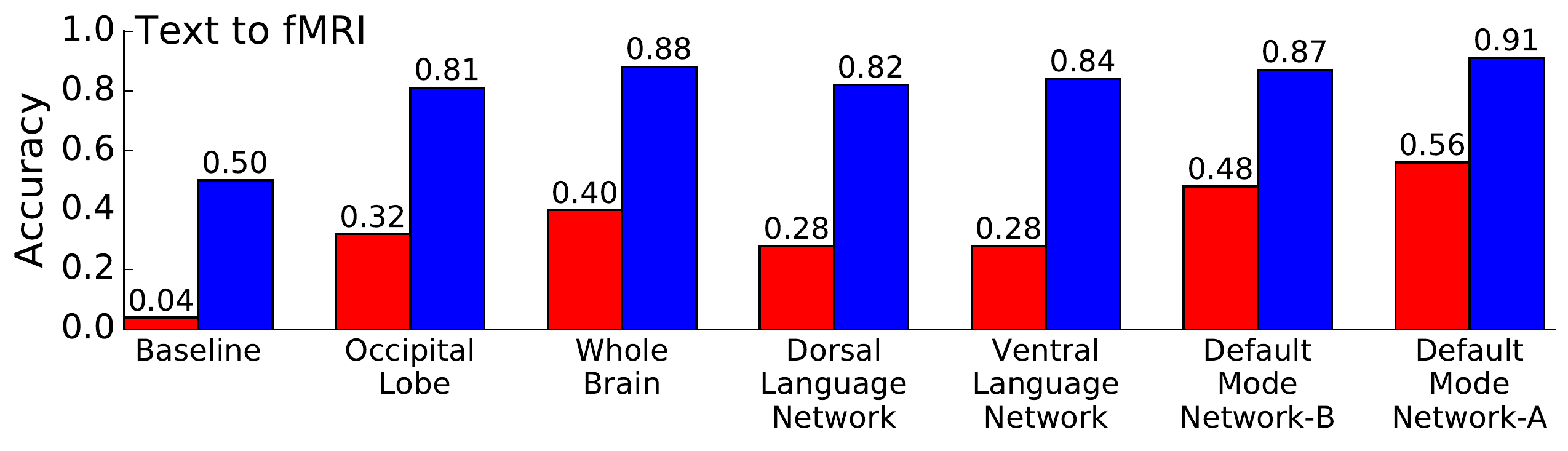}
\caption{Best Bidirectional Accuracy Scores for Each Brain Region of Interest for both Scene Classification and Ranking: In this figure, for each ROI and for each experiment (Text $\to$ fMRI $4\%$ (red), $50\%$ (blue) chance rates; fMRI $\to$ Text $4\%$ (red), $50\%$ (blue) chance rates), we give the best performance as a percentage. For all measures, closer to $100\%$ is better. We can see that Whole Brain, DMN-A, and DMN-B tend to perform the best, and that fMRI $\to$ Text performs better than Text $\to$ fMRI.}
\label{figure:topROIacc}
\end{figure}

\subsection{Top Absolute Performances over All Algorithms}

Figure \ref{figure:topROIacc} demonstrates that the DMN regions have nearly the best performance over the other ROIs studied, which fits with prior research in this area (\cite{Regev13}, \cite{Simony16}). We achieve $72\%$ accuracy over $4\%$ chance with the Whole Brain region in the scene classification task. Since the scene ranking measure is always $\geq 80\%$, the average rank of the correct answer is in the top $20\%$ of the scenes, which translates to top $5$ scenes out of $25$. For fMRI $\to$ Text we perform even better, where the average rank of the correct answer is in the top $10\%$ of the scenes (top $3$ scenes out of $25$). Notably, we get excellent performance out of the Whole Brain region, which has $26000$ voxels selected by merely choosing voxels whose intersubject correlation is above a certain threshold. This result demonstrates that our methods are not overly dependent on applying domain-specific knowledge (we do not necessarily have to preselect an ROI to get good results).

\textbf{fMRI $\to$ Text}. Here we discuss the performance of the fMRI $\to$ Text experiments. In Figure \ref{figure:topROIacc}, we display the top accuracy over all algorithmic choices for each experiment. We achieve high accuracy performance, reaching $72\%$ for the scene classification task for fMRI $\to$ Text and in the mid-$90\%$s for the scene ranking tasks. In particular, the Whole Brain and the DMN regions perform best, supporting previous work by \cite{Regev13} and others demonstrating that the DMN plays an important role in narrative processing.

\textbf{Text $\to$ fMRI}. On the other hand, we see that the Text $\to$ fMRI experiments perform worse than the fMRI $\to$ Text experiments. The best top$-1$ scene classification accuracy performance is $56\%$ for the DMN-A region, and the other top performing regions get accuracy in the mid-to-high $40\%$ accuracy. For the ranking task, performance ranges from $80\%-90\%$, which is again slightly worse than the fMRI $\to$ Text ranking experiment.

\subsection{Comparing Algorithmic Choices}

\begin{figure}[!t]
\centering
\makebox[\textwidth][c]{\includegraphics[width=1.5\textwidth]{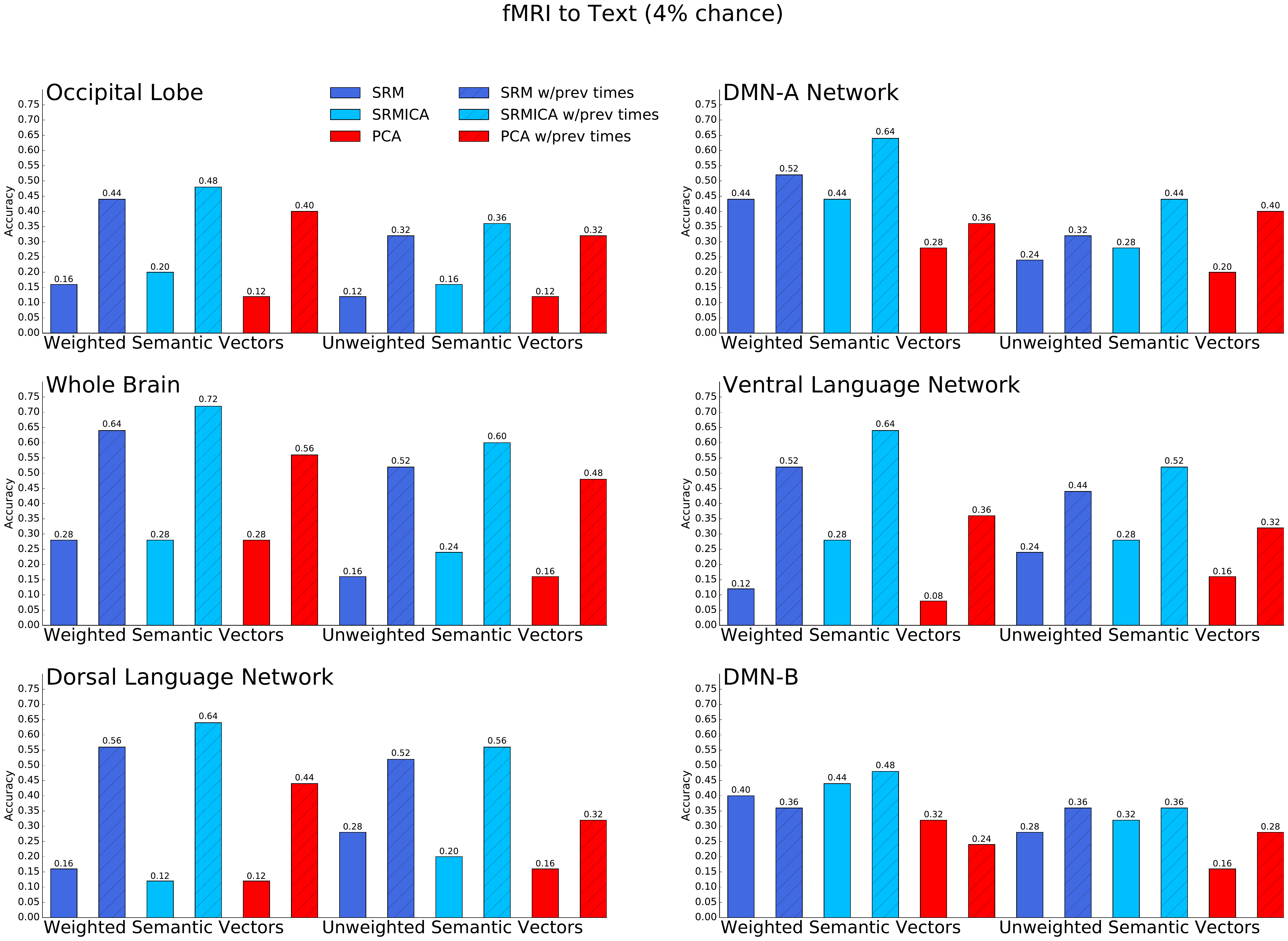}}
\caption{Comparisons for all ROIs for the fMRI $\to$ Text Top-$1$ Scene Classification Experiment: The chance rate for this task is $4\%$. Each plot is for a different ROI. Here, we only display results which use the Procrustes linear map since it on average performs better than ridge regression for fMRI $\to$ Text. We also fix the number of previous time points used for the shaded bars at $8$ previous time steps, since that tends to be near optimal. We present comparisons between SRM/SRM-ICA and PCA using blue colors versus red colors, and compare weighted semantic aggregation (left) to unweighted semantic aggregation (right) by x-axis position.}
\label{figure:FT4}
\end{figure}

\begin{figure}[!t]
\centering
\makebox[\textwidth][c]{\includegraphics[width=1.5\textwidth]{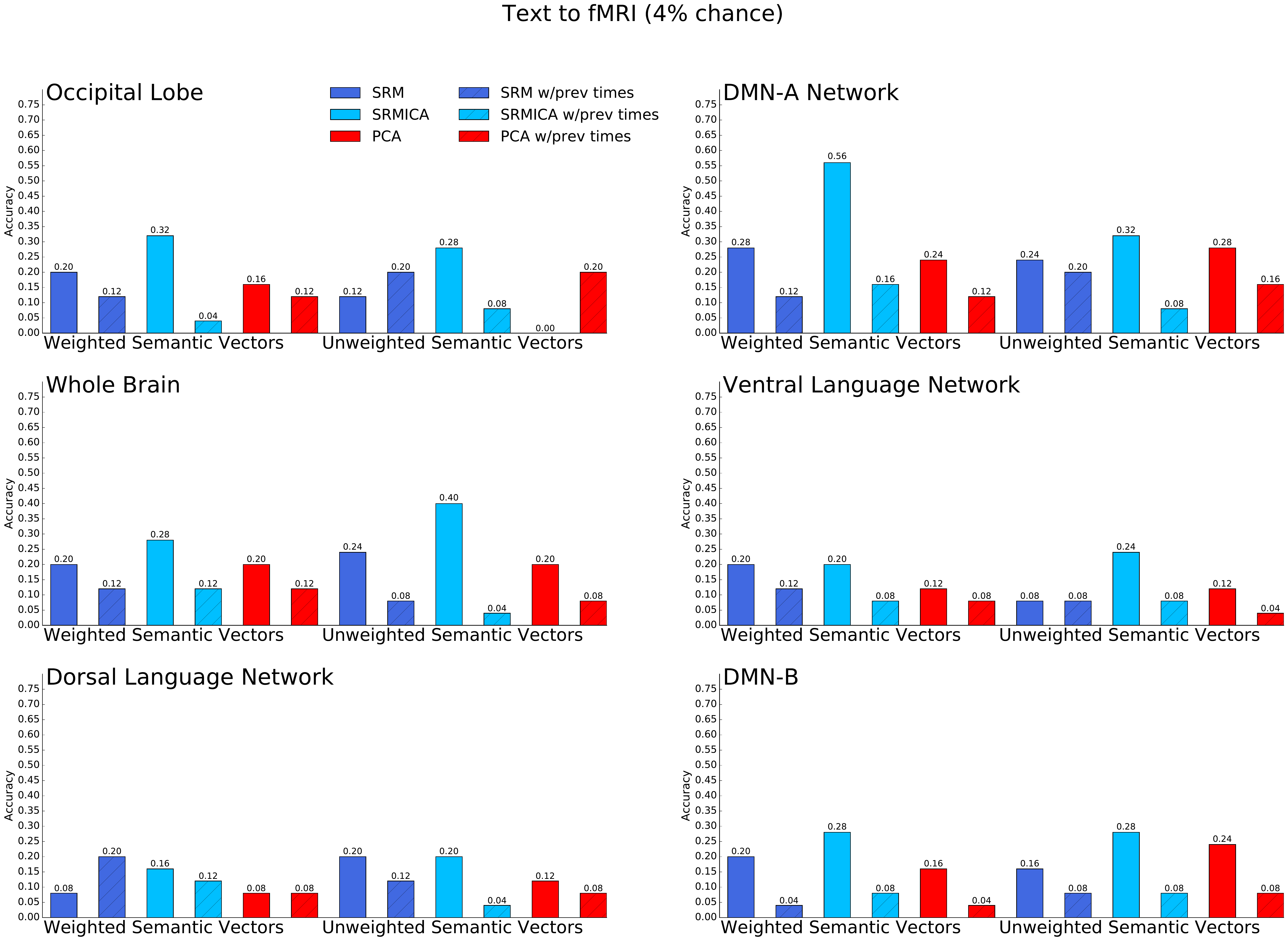}}
\caption{Comparisons for all ROIs for the Text $\to$ fMRI Top-$1$ Scene Classification Experiment: The chance rate for this task is $4\%$. Each plot is for a different ROI. Here, we only display results which use the ridge regression linear map since it on average performs better than Procrustes for Text $\to$ fMRI. We also fix the number of previous time points used for the shaded bars at $8$ previous time steps, since that tends to be near optimal. We present comparisons between SRM/SRM-ICA and PCA using blue colors versus red colors, and compare weighted semantic aggregation (left) to unweighted semantic aggregation (right) by x-axis position.}
\label{figure:TF4}
\end{figure}

In order to simplify presentation for Figures \ref{figure:FT4} and \ref{figure:TF4}, we chose to fix the algorithmic parameters that uniformly outperformed other options. All linear maps for fMRI $\to$ Text were learned using the Procrustes method and all linear maps for Text $\to$ fMRI were learned using the ridge regression approach. We fixed these for comparison purposes since, for fMRI $\to$ Text scene classification, Procrustes performed $1.25\times$ better than ridge on average (Table \ref{table:avg_deltas}). On the other hand, ridge performed $1.2\times$ better than Procrustes on average over Text $\to$ fMRI scene classification (Table \ref{table:avg_deltas}). As a caveat, there were exceptions to the rule, as the max ratios in Table \ref{table:avg_deltas} indicate. In Figures \ref{figure:FT4} and \ref{figure:TF4}, for the data points that are labeled as using previous time steps, we reported the result for $8$ previous time steps. The optimal number of previous time steps for fMRI $\to$ Text was typically between $5-8$, and so we fixed that choice of parameter across all of the graphs in these figures.

\vspace{1.5em}

\textbf{Comparing SRM and SRM-ICA to PCA}. We see considerable improvement on best-case performance when using SRM or SRM-ICA over PCA, particularly on the fMRI $\to$ Text tasks, in some cases gaining as much as $1.8\times$ the top$-1$ scene classification performance of PCA, as demonstrated in Figure \ref{figure:FT4}. Typically, SRM-ICA tends to perform slightly better, especially on the Whole Brain ROI. The case is weaker for Text $\to$ fMRI, since though we can find that performance increases by as much as $2.3\times$ the top$-1$ scene classification performance, the average benefit is smaller (Table \ref{table:avg_deltas}, Figure \ref{figure:TF4}). If we look at average case improvements, we see considerable gains in both directions: SRM/SRM-ICA improve on average by $1.3\times$ over PCA for fMRI $\to$ Text scene classification, and on average by $1.2\times$ over PCA on Text $\to$ fMRI scene classification. For the ranking tasks, we note that while performance improvement for the best selections of algorithm parameters is not as distinct, SRM and SRM-ICA can drastically improve upon PCA performance for poor selection of parameters. This fact suggests that one should always use SRM or SRM-ICA over PCA, since on new datasets where it is not known which linear map to use, or the number of previous time points to incorporate in the analysis and so on, our results here suggest that these SRM-variants will improve strongly upon PCA if the parameters are poorly chosen, and still improve decently upon PCA otherwise.  

\vspace{1.5em}

\textbf{Weighted vs. Unweighted Aggregation of Word Embeddings}. Using the SIF-weighted embeddings improves upon unweighted averaging when featurizing the annotation vectors as well. Examining Table \ref{table:avg_deltas} and Figure \ref{figure:FT4}, we see that for fMRI $\to$ Text top$-1$, there is improvement on best-case performance by as much as $1.3\times$ by using weighted embeddings. On average, we see that weighted embeddings improve by $1.2\times$ over the unweighted embeddings. Looking at Figure \ref{figure:TF4}, the case is weaker for Text $\to$ fMRI top$-1$; while for some algorithms and ROIs we see as much as $2.5\times$ improvement on best-case performance by weighted aggregation embeddings, we also see that sometimes unweighted averaging can outperform weighted averaging. However, on average, weighted embeddings improve by $1.1\times$ over unweighted averaged embeddings.

\vspace{1.5em}

\begin{figure}[!t]
\centering
\makebox[\textwidth][c]{\includegraphics[width=1\textwidth]{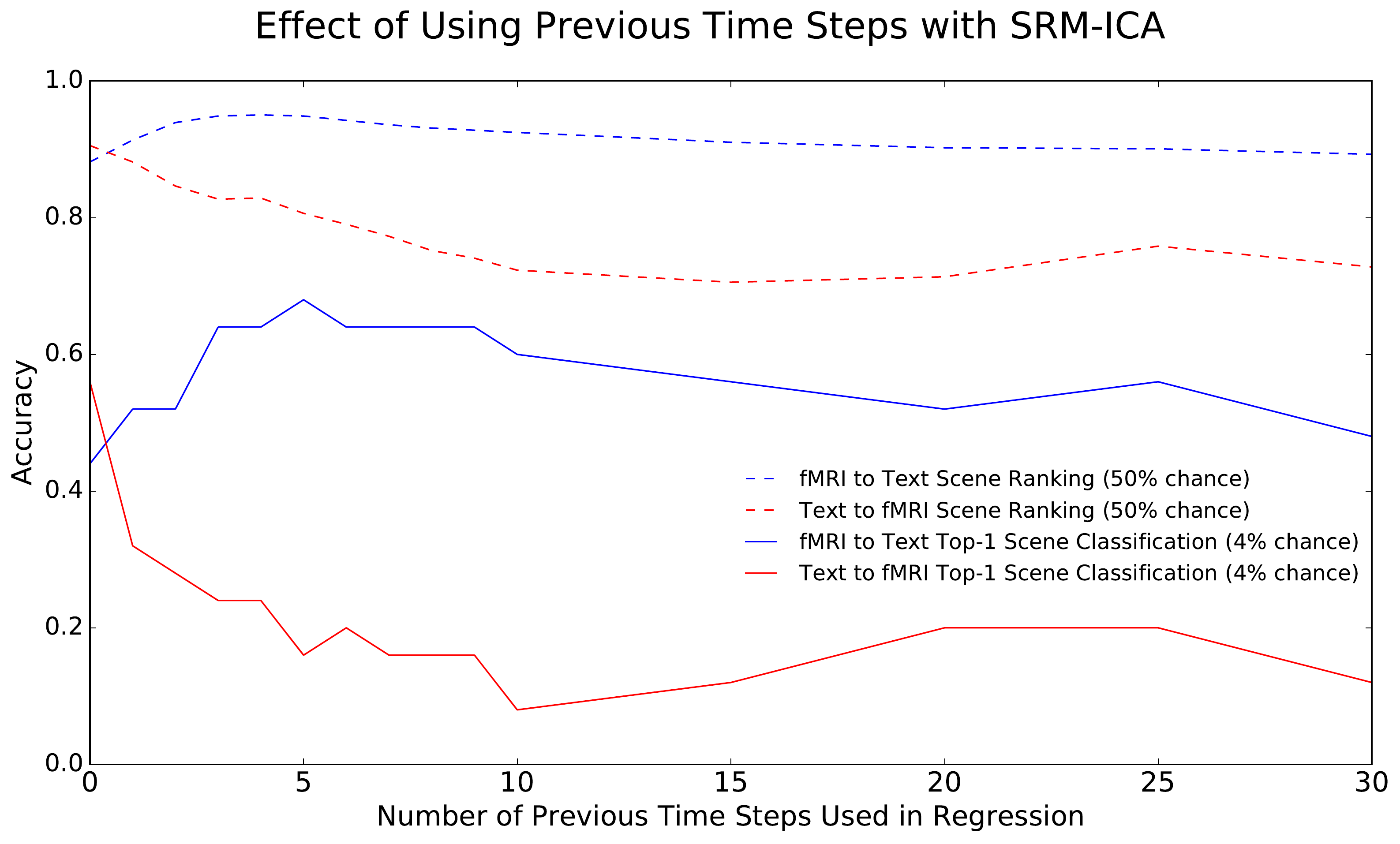}}
\vspace{-2.5em}
\caption{Varying Previous Timesteps: For the DMN-A region, choosing SRM-ICA, weighted average, Procrustes for the fMRI $\to$ Text linear map, and ridge for the Text $\to$ fMRI linear map, we plot the relationship between accuracy (y-axis) and number of previous time points used in the linear map fit (x-axis). We can see a peak at around using $5-8$ previous TRs as optimal for the fMRI $\to$ Text tasks, and a relatively monotone decay for using any previous TRs in the Text $\to$ fMRI tasks.}
\label{figure:prev_timepoints}
\end{figure}

\textbf{The Effects of Previous Time Points}. Figure \ref{figure:FT4} demonstrates the positive effect of adding previous time steps to the accuracy scores for the fMRI $\to$ Text case. Table \ref{table:avg_deltas} demonstrates that at best, using previous timepoints can improve performance by as much as $5.3\times$. On average, this improvement is $1.8\times$, nearly doubling performance. On the other hand, Figure \ref{figure:TF4} shows that for Text $\to$ fMRI, adding previous time steps almost universally hurts performance and on average halves performance (Table \ref{table:avg_deltas}). This fact is also evident from Figure \ref{figure:prev_timepoints}, which illustrates the situation for the DMN-A ROI.  

Notably, the effect of using previous time steps is different from learning a hemodynamic response function, which other authors \cite{Wehbe14, Huth16} have done in the past. Instead, we are investigating whether information from longer time scales helps improve performance. In Figure \ref{figure:prev_timepoints}, we see that there are some peaks in classification performance between $5$ and $8$ previous time steps ago (or $7.5 - 9.0$ seconds ago, after having taken into account the HRF). However, using any number of previous time steps (up to as long as $30$ TRs ago, or $45$ seconds) still improves over the baseline of using no previous time steps. 

For Text $\to$ fMRI however, the story is different. We see no improvement in performance when using previous time points, and in fact performance decreases (Figure \ref{figure:prev_timepoints}). A possible explanation for this result is the substantially higher dimensionality of the text annotation embeddings ($100$ dimensions) compared to the fMRI embeddings ($20$ dimensions). Thus, adding text vectors from previous timesteps in this ``stacked'' fashion greatly increases model complexity and may therefore contribute to significant overfitting and poor generalization performance.

\vspace{-1em}
\section{Conclusion}

\vspace{-0.5em}

In this paper, we have explored several methods that improve our success at mapping between fMRI response to a natural stimulus and semantic text data describing this stimulus. We see that SRM and SRM-ICA perform considerably better than simple averaging or using PCA. Figure \ref{figure:FT4} demonstrates that weighted aggregation of the words in semantic space to form annotation vectors over simple averaging improves the baseline accuracy by a reasonable amount. We also show that adding previous time steps improves accuracy substantially.

Using SRM-ICA in fMRI space, weighted annotation vectors in semantic space and a Procrustes linear map learned between the concatenations of five previous time points in fMRI and semantic space, we are able to achieve $72\%$ scene classification accuracy over $4\%$ chance rate for the Whole Brain region on the fMRI $\to$ Text task. 

Other ROIs are typically above $60\%$ scene classification accuracy as well. Similarly, in the scene ranking task, we achieve $> 90\%$ average rank for the correct answer across ROIs. Text $\to$ fMRI does not perform as well but is still far above chance ($56\%$ with DMN-A ROI for $4\%$ chance rate, and $> 80\%$ average rank across ROIs). Another takeaway is that SRM and SRM-ICA improve upon PCA almost always, and provide particularly substantial improvement in cases where the other parameter settings (like the semantic featurization or selection of linear map and associated hyper-parameters) are not necessarily tuned. These results indicate that we are able to use multiple subjects to learn a $20$-dimensional shared space for the fMRI data that increases performance on our experiments. Thus, we provide concrete evidence towards the hypothesis made in \cite{Huth16} regarding the existence of a \textbf{shared} fMRI representation across multiple subjects that correlates significantly with \textbf{fine-grained} semantic context vectors derived via statistical word co-occurrence properties. 

The method of combining word vectors is another essential part of our results. 
We demonstrate that weighted-SIF averaging \cite{Arora17} for aggregating individual elements of a word sequence performs on average $1.2\times$ better than unweighted averaging for fMRI $\to$ Text top$-1$ scene classification, and on average $1.1\times$ better for Text $\to$ fMRI top$-1$ scene classification. Since we use only semantic vectors to featurize a movie stimulus dataset, our work provides additional support for the notion that the distributional hypothesis of word meaning may extend to real life multi-sensory stimuli.

Finally, we note that using multiple previous timepoints when mapping from fMRI $\to$ Text is very beneficial and significantly improves results by a factor of as much as $5.3\times$, and on average nearly doubles performance (Table \ref{table:avg_deltas}). 

We did not see a benefit of using previous timepoints on mapping from Text $\to$ fMRI, but -- as stated above -- this may be an artifact of our ``stacking'' procedure, which greatly increases the dimensionality of the input when using previous Text timepoints. In future work, we plan to explore whether other approaches to including previous timepoints (e.g., learning a weighted average) yield better results. 

We also plan to investigate other datasets to see if our techniques in matching fMRI and Text generalize to more varied stories, in both audio and visual formats.

\section{Acknowledgments}

The dataset is online \cite{Chen17} and the code used in this paper will be made available on GitHub. Additionally, we note that we used \texttt{http://brainiak.org/} for some of the implementations of algorithms used in this paper. This work was funded by a grant from the Intel Corporation, NIMH R01MH112357 awarded to U. Hasson and K. Norman; NIH grants R01-MH094480 and 2T32MH065214-11; NSF grants CCF-1527371, DMS-1317308, Simons Investigator Award, Simons Collaboration Grant, and ONRN00014- 16-1-2329 awarded to S. Arora, and NSERC Discovery Grant RGPIN 2014-04465 awarded to C. Honey. P. H. Chen was supported by a Google Fellowship. 

%
% ---- Bibliography ----
%
\bibliography{neuro_image_fMRI_Text_KAN_FINAL}
\bibliographystyle{splncs}

\end{document}